\documentclass[doublespacing]{elsart}

\usepackage{graphics}
\usepackage{graphicx}
\usepackage{epsfig}

\usepackage{amssymb}
\def\lsim{\raise0.3ex\hbox{$<$\kern-0.75em\raise-1.1ex\hbox{$\sim$}}}
\def\gsim{\raise0.3ex\hbox{$>$\kern-0.75em\raise-1.1ex\hbox{$\sim$}}}

\newcommand{\be}{\begin{equation}}
\newcommand{\ee}{\end{equation}}

\def\beq{\begin{equation}}
\def\eeq{\end{equation}}
\def\beqa{\begin{eqnarray}}
\def\eeqa{\end{eqnarray}}

\def\gappeq{\mathrel{\rlap {\raise.5ex\hbox{$>$}}
{\lower.5ex\hbox{$\sim$}}}}

\def\lappeq{\mathrel{\rlap{\raise.5ex\hbox{$<$}}
{\lower.5ex\hbox{$\sim$}}}}

\def\Toprel#1\over#2{\mathrel{\mathop{#2}\limits^{#1}}}

\begin{document}

\begin{frontmatter}

\title{Looking for intrinsic charm in the forward region at BNL RHIC and CERN LHC}

\author[Victor]{V. P. Gon\c calves\corauthref{cor1} }
\ead{barros@ufpel.edu.br}
\corauth[cor1]{Corresponding author.  Phone/Fax: 55 53 32757477.}
\address[Victor]{Instituto de F\'{\i}sica e Matem\'atica,\\
Universidade Federal de Pelotas,
Caixa Postal 354, CEP 96010-900, Pelotas, RS, Brazil}
\author[navarra]{F. S. Navarra}
\ead{navarra@if.usp.br}
\address[navarra]{Instituto de F\'{\i}sica, Universidade de S\~{a}o Paulo,
C.P. 66318,  CEP 05315-970, S\~{a}o Paulo, SP, Brazil}
\author[thomas]{T. Ullrich}
\address[thomas]{Physics Department, Brookhaven National Laboratory\\ 
Upton, New York, 11973, USA}

\begin{abstract}
The complete understanding of the basic constituents of  hadrons and the hadronic dynamics 
at high energies are two of the main challenges for the theory of  strong interactions. In 
particular, the existence of  intrinsic heavy quark components in the hadron wave function 
must be confirmed (or disproved). In this paper we propose a new mechanism for the production  
of $D$-mesons at forward rapidities based on the Color Glass Condensate (CGC) formalism and 
demonstrate that the resulting transverse momentum spectra are strongly dependent on the 
behavior of the charm distribution at large Bjorken $x$. Our results show clearly that the 
hypothesis of  intrinsic charm can be tested in $pp$ and $p(d) A$ collisions at RHIC and LHC.
\end{abstract}

\end{frontmatter}
\section{Introduction}
\label{intro}
A complete knowledge of the partonic structure of the hadrons is fundamental to make predictions for both the Standard Model and beyond Standard Model processes at hadron colliders.  Since the early days of the parton model and of the first deep inelastic scattering  (DIS)  experiments, determining the precise form of the quark and gluon distributions of the nucleon has been   a major goal of high energy hadron physics. Over the last 40 years enormous progress has  been achieved. In particular, data from HERA have dramatically improved our knowledge about the small-$x$ behavior of the parton distributions functions (PDF's), finding a strong rise of the structure functions in this kinematical range, which implies a high partonic density in the proton. From  theoretical considerations it is expected that  in the high energy limit the small-$x$ gluons in a hadron wavefunction should form a Color Glass Condensate (CGC). The current experimental results at HERA and 
RHIC  provide some  evidence for the CGC physics (For a review see, e.g., \cite{cgc}). 
Although they are not yet conclusive, these experiments show  the manifestation 
of non-linear (saturation)  effects in  high energy processes. The measured 
multiplicities and rapidity distributions in heavy ion collisions, the high $p_T$
suppression of particle production at large rapidities in deuteron-gold collisions,  
geometric scaling and other observations \cite{cgc}  point towards saturation. 
Of course, so far one has been  able to explain these measurements with other 
approaches. For example, 
in Ref. \cite{forte} it was shown that geometric scaling can be derived from the DGLAP 
equations \cite{dglap}. From our
point of view, the most important consequence of these observations  is that a change 
of formalism seems to be required. In particular it seems to be necessary to replace 
the collinear factorization formula by a more powerful formalism, able to accommodate  
saturation effects if and when they are present.
 
Another important improvement has occurred in our knowledge about the heavy quark contribution 
to the proton structure. In the last years several groups have proposed different schemes to 
determine these distributions considering that  the heavy quark component in PDF's can be 
perturbatively generated by gluon splitting (See e.g. \cite{tung}). This component is usually 
denoted  {\sl extrinsic} heavy quark  component. Moreover, the possibility of an {\sl intrinsic}
  component has been studied in detail and  included in the recent versions of the CTEQ 
parameterization \cite{cteq}. The hypothesis of intrinsic heavy quarks (IHQ) is a natural 
consequence of the quantum fluctuations inherent to Quantum Chromodynamics (QCD) and amounts to  
assuming the existence of a $Q\bar{Q}$ ($Q = c,b,t$) as a nonperturbative component in the 
hadron wave function.  One  of the most striking properties of an IHQ state, such as 
$|uudQ\bar{Q}\rangle$, is that the heavy constituents tend to carry the largest fraction of the 
momentum of the hadron. Consequently,  in 
contrast to heavy quarks produced through  usual perturbative QCD, which emerge  
with small longitudinal momentum,  the intrinsic charm component gives rise to heavy mesons  
with large fractional momenta relative to the beam particles. Therefore, 
the existence of an intrinsic component modifies, for instance, the $x_F$ and rapidity 
distribution of charmed particles (See, e.g. \cite{ingelman}). Moreover, it can also lead 
to Higgs production at high $x_F$ \cite{brod_higgs}. Many of these features have been 
discussed in the pioneering works on intrinsic charm \cite{bhps,vb,hal}. 
 
 Although the existence of IHQ fluctuations in the proton has substantial and growing experimental and theoretical support, as revised in  Ref. \cite{pumplin} and briefly discussed below, more definite conclusions are still not possible. One of the shortcomings of previous studies  is  the use of the QCD factorization at large Feynman $x_F$, where it is expected to breakdown (See, e.g., \cite{kop_fac}). Basically, at large $x_F$ the kinematics is very asymmetric, with the hadrons in the final state emerging from collisions of projectile partons with large light cone momentum fractions ($x_p \rightarrow 1$) with target partons carrying a very small momentum fraction 
($x_t \ll 1$). In this case, we have a scattering of a dilute projectile on a dense target, where the small-$x$ effects coming from CGC physics are expected to show up and the usual factorization formalism is expected to breakdown \cite{raju_fac}. This kinematical region has been explored in dAu collisions at forward rapidities at RHIC \cite{BRAHMSdata} and the suppression of high $p_T$ hadron yields,  anticipated on the basis of CGC ideas, was observed. The satisfactory description of these experimental data \cite{buw,betemps} gives us a strong indication that the appropriate framework to calculate  particle production at large $x_F$ (large rapidities) is one based on CGC physics. Having this in mind, in this paper we adapt the formalism proposed in \cite{dhj} 
(to calculate the production of  high-$x$ quarks in $dAu$ collisions) to $D$ meson production 
in $pp$ ($p(d)A$) collisions at forward rapidities, including  intrinsic charm quarks in the 
projectile wave function. In 
this approach the target is treated as a dense system  while the projectile proton is taken to be a dilute system in the spirit of standard QCD. We shall assume that the charm 
quarks are already in the projectile and their densities are those given by the CTEQ 
parameterization \cite{cteq}. Moreover, we will describe the interaction of these charm quarks in terms of the dipole scattering amplitude recently proposed in \cite{buw} to describe the RHIC data. Finally, 
we will assume that the charm quarks fragment into $D$ mesons according to the fragmentation 
functions given in Ref. \cite{bkk}. Our goal is to estimate the $D$-meson spectrum at forward 
rapidities at RHIC and LHC energies assuming the presence (or absence) of an intrinsic 
charm component and evaluate how much this quantity is modified. We emphasize that in our 
calculation  there is no new free parameter. 

In what follows we show that if an intrinsic charm component is present in the projectile wave 
function, it will dominate the $D$-meson production at forward rapidities, being more important 
than the other possible mechanisms. The paper is structured as following. In the next section 
(Sec. \ref{sec2}) we present a brief review of the intrinsic charm models and discuss the 
predictions of the CTEQ collaboration for the $x$ dependence of the charm distribution. 
In Sec. \ref{sec3} we discuss  charm production at forward rapidities   
and propose a new mechanism based on the CGC formalism. In Sec. \ref{sec4} our predictions for the transverse momentum distribution of $D$-mesons produced in $pp$ and $p(d)A$ collisions at RHIC and LHC energies are presented. Finally, in Sec. \ref{sec5} we summarize our main conclusions.

\section{Intrinsic Charm Models}
\label{sec2}
The charm content of the nucleon sea comes mostly from the DGLAP  
evolution of the initial gluon distribution. This process is well understood 
in perturbative QCD.  However, as pointed out long ago in Ref. \cite{bhps} (see   
also Ref. \cite{hal})   there may be another, 
higher twist, component which is typical of the hadron where it is created. 
This component is called intrinsic charm (IC). A comprehensive review of the main 
characteristics of the  IC models can be found in \cite{pumplin}.  A well known 
model was proposed in Ref. \cite{vb}. In this model, the creation of the 
$c \overline{c}$ pair was studied in detail. It was assumed that the nucleon 
light cone wave function has  higher Fock states, the first one being 
$|q q q c \overline{c}>$. The probability of finding the nucleon in this 
configuration was given by the inverse of the squared invariant mass of the 
system. Because of the heavy charm mass, this probability as a function of the 
quark fractional momentum, $P(x)$, is very hard, as compared to the one obtained 
through the DGLAP evolution. We denote this model by BHPS.

A more dynamical perspective is given by the meson 
cloud model. In this model, the nucleon fluctuates into an intermediate state  
composed by a charmed baryon plus a charmed meson \cite{pnndb}. 
The charm is always confined in one hadron and 
carries the largest part of its momentum. In the hadronic description we can use 
effective lagrangians to compute the charm splitting functions, which turn out to 
favor  harder charm quarks than the DGLAP ones. 
The main difference between the BHPS and meson cloud models is that the latter predicts that  
the charm and anticharm distributions are different, which generates  a difference in the  
momentum distributions of $D^+$ and $D^-$ \cite{cdnn}.

In spite of its successes, the idea of intrinsic charm was left aside for some years.  
Very recently it  was considered again as 
an ingredient in the global fit of DIS data performed by the CTEQ collaboration 
\cite{cteq}. In this update the CTEQ group determined the shape and normalization 
of the IC distribution in the same way as they do for other parton species. 
In fact they find several IC distributions which are compatible with the world data. Apart from 
the already mentioned BHPS and meson cloud models, the CTEQ group has tested another model of
intrinsic charm, called sea-like IC. It consists basically in assuming that at a very low 
resolution (before the DGLAP evolution) there is already some charm in the nucleon, which has a 
typical sea quark momentum distribution ($\simeq 1/ \sqrt{x}$) with normalization to be fixed by 
fitting DIS data.

The different parameterizations are presented  in Fig. \ref{fig1}. For comparison we also present the no-IC distribution, where the charm content of the nucleon sea comes from the DGLAP evolution (exclusive charm component).  We can see that the 
intrinsic charm distribution is a factor ten (sea like) to twenty (BHPS) larger than the no-IC distribution 
(the long dashed line)  and the peaks of the IC distributions lie in the large $x$ 
($> 0.1$) region. This behavior is so peculiar  that it gives us hope to observe IC 
experimentally.

\section{$D$-meson production at forward rapidities in the CGC formalism}
\label{sec3}
Heavy quark production at high energies is usually described in the collinear or $k_T$-factorization frameworks of QCD. Using the collinear factorization, charm production  is described  in terms  of the  basic subprocesses of gluon fusion ($ g + g \rightarrow c + \overline{c}$) and light quark-antiquark fusion   ($ q + \overline{q} \rightarrow c + \overline{c}$). The elementary cross section computed to leading order (LO) or next-to-leading order (NLO) is folded with the corresponding parton distributions and fragmentation functions. This is the basic procedure in most of the calculations performed, for instance,  with the standard code PYTHIA. However, at small-$x$ the collinear factorization should be generalized to resum powers of $\alpha_s \ln (s/q_T^2)$, where $q_T$ is the transverse momentum of the final state and $\sqrt{s}$ is the center of mass energy. 
This resummation is done in the $k_T$-factorization framework, where the cross section is expressed in terms of unintegrated gluon distributions which are determined by the QCD dynamics at small-$x$ (For a review see e.g.  \cite{smallx}). 
In Ref. \cite{kt_hq} the authors have estimated the heavy quark production using the $k_T$-factorization scheme assuming the dominance of the gluon-gluon interactions and the presence of high density effects and predicted a much harder open charm spectrum. More detailed studies, considering for instance the breakdown of the $k_T$-factorization,   have been performed in Refs. \cite{hq_sat}.
In general, these studies assume the dominance of gluon interactions and this implies a strong decrease of the spectrum at large rapidities, since in this regime the cross section is dominated by very large $x_p$ and small values of $x_t$. 
From light hadron production, we know  that in this kinematical range the cross section is 
dominated by the interaction of valence quarks of the projectile. In other words, the cross section is dependent on the partonic structure of the projectile at large-$x$. Therefore, the basic assumption of the previous studies about heavy quark production at high energies and forward rapidities is satisfactory if an intrinsic heavy quark component in the proton wave function is not considered. On the other hand, if this component is present and strongly modifies  the behavior of the charm parton distribution at large $x$, as shown in Fig. \ref{fig1}, one must estimate its contribution to the $D$-meson production cross section.

In order to calculate the intrinsic charm contribution we generalize the 
DGLAP $\otimes$ CGC factorization scheme proposed in Ref. \cite{dhj}, which describes 
quite well the experimental data on light hadron production at forward rapidities in 
$pp/dA$ collisions \cite{buw} (See also \cite{betemps}). In this approach the projectile (dilute system) evolves 
according to the linear DGLAP dynamics and the target (dense system) is treated using 
the CGC formalism. Apart from \cite{dhj} we refer the reader to \cite{jamaldumi} for 
the first derivation of the main formula used here and to \cite{kt_hq} and 
\cite{hq_sat} for other applications of this formalism, including open charm production
at RHIC (without intrinsic charm component).  It is important to emphasize that, as shown 
in \cite{dhj}, this approach correctly reproduces the results of the linear regime of QCD 
and collinear factorization in the low density regime.

In what follows,  we assume the existence of charm quarks with large $x$ inside the 
proton, which interact with  the target and 
subsequently fragment into  $D$-mesons. This implies that  single-inclusive charm 
production in hadron-hadron (nucleus) processes will be  described in the CGC formalism 
by 
\begin{eqnarray}
&\,& {d N \over dy \, d^2 p_T }(p p(A) \rightarrow D X) = 
{1 \over (2\pi)^2}
\int_{x_F}^{1} dx_p \, {x_p\over x_F} 
f_{c/p} (x_p,Q^2)~ \nonumber \\
&\,& \times {\cal{N}_F} \left(x_2,{x_p\over x_F}p_T \right)~ D_{D/c} 
\left({x_F\over  x_p}, Q^2\right) ,
\label{secdif}
\end{eqnarray}
where $p_T$, $y$ and $x_F$ are the transverse momentum, rapidity  and the Feynman-$x$
of the produced D meson, respectively. The variable $x_p$ denotes the momentum
fraction of the charm inside the projectile.
Moreover,  $f_{c/p}(x_p,Q^2)$ is the projectile charm 
distribution function, taken from the CTEQ group,  and $D(z, Q^2)$ the charm 
fragmentation function into $D$ mesons taken from Ref. \cite{bkk}.  
The factorization scale is chosen to be $Q^2 = p^2_T$. These quantities  evolve according to 
the DGLAP evolution equations and respect the momentum
sum-rule. In Eq. (\ref{secdif}), ${\cal{N}_F}(x,p_T)$  is  the fundamental  
representation of the  forward dipole amplitude in  momentum space determined by 
CGC physics. As explained below, it is obtained from ${\cal{N}_A}(x,k_T)$, 
the forward dipole amplitude in the adjoint representation, which is given by: 
\begin{eqnarray}
&  & {\cal{N}}_A(\vec{q}_t,x_2)  \equiv   \int d^2 r_t  e^{i \vec{q}_t
    \cdot \vec{r}_t} {\cal{N}}_A(\vec{r}_t,\vec{q}_t,x_2)\nonumber\\
& = &  - \int d^2 r_t  e^{i \vec{q}_t
    \cdot \vec{r}_t} \left[1-\exp\left(-\frac{1}{4}(r_t^2
      Q_s^2(x_2))^{\gamma(\vec{q}_t,x_2)}\right) \right]  \nonumber \\
\label{NA_param}
\end{eqnarray}
where $r_t$ is the transverse size of the dipole and  $Q_s$ is the saturation scale 
given by $Q_s(x) = 1\,\mathrm{GeV}\,
\left(\frac{x_0}{x}\right)^{\lambda/2}$
with $x_0 \simeq 3 \times 10^{-4}$ and $\lambda \simeq 0.3$. For nuclear
targets  $\, Q_s^2$ contains an additional factor $A^{1/3}$.
The function $\gamma$ in (\ref{NA_param}) is the anomalous dimension, which encodes 
the physical content of the dipole amplitude. We follow \cite{buw} and parametrize it
as: 
\begin{equation}
 \gamma(w)=\gamma_1+(1-\gamma_1)\frac{(w^a-1)}{(w^a-1)+b}\,.
\label{gamma_alpha}
\end{equation}
where   $w=q_t/Q_s(x_2)$ and $\gamma_1 = \gamma(w=1)$. 
The two free parameters $a$ and $b$ were  fitted to the RHIC data \cite{buw}.
The corresponding expression ${\cal{N}}_F$ for quarks is obtained from ${\cal{N}}_A$
by the replacement $(r_t^2Q_s^2)^\gamma \to ((C_F/C_A) r_t^2Q_s^2)^\gamma$,  
with $C_F/C_A=4/9$. No $K$ factor (to account for higher order
corrections) is introduced. The basic characteristic of this model for the forward dipole amplitude  is that it explicitly satisfies the geometric scaling property. Besides, in this model the large $p_T$ limit, $\gamma \rightarrow 1$, is approached much faster in comparison with other phenomenological parameterizations, which implies different predictions for the large $p_T$ slope of the hadron and photon yield (For a detailed discussion see Ref. \cite{betemps}). Finally, we would like to emphasize that this saturation model is able to describe the experimental data for the light hadron production at mid- and forward rapidities.

A comment is in order. In a full calculation of   
$D$ production in the CGC formalism we should consider the IC contribution estimated in this paper as 
well as the contribution which comes from gluon-gluon interactions, as estimated for 
instance in Refs. \cite{kt_hq,hq_sat}. However, as these contributions strongly decrease at large rapidities, we believe that our main conclusions are not modified by the inclusion of these terms.
 
\section{Results}
\label{sec4}
\begin{figure}[t]
\includegraphics[scale=0.55]{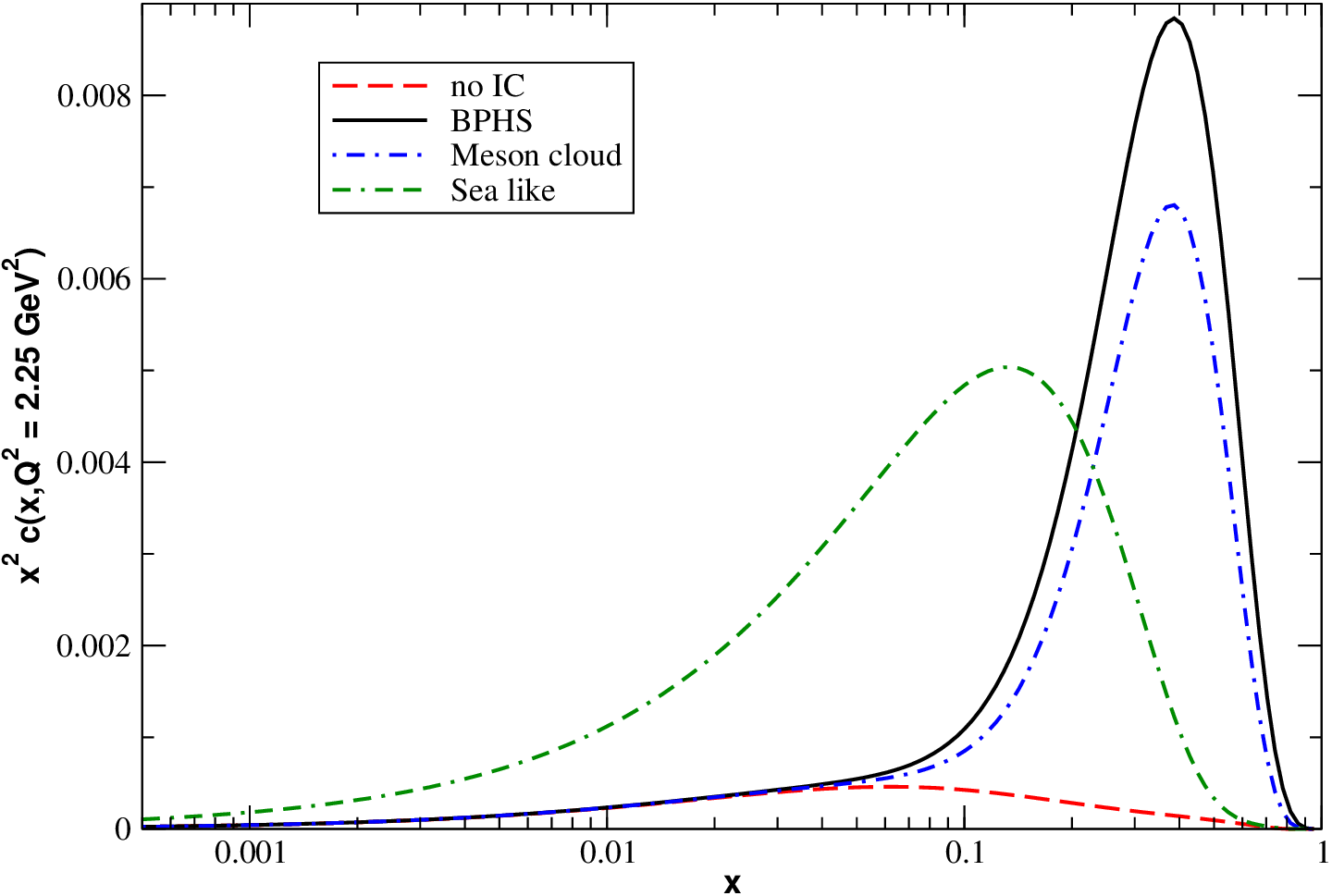}
\caption{Intrinsic charm momentum distribution in the proton  with different IC models
taken from \cite{cteq}.} 
\label{fig1}
\end{figure}

\begin{figure}[t]
\begin{tabular}{c}
\includegraphics[scale=0.45] {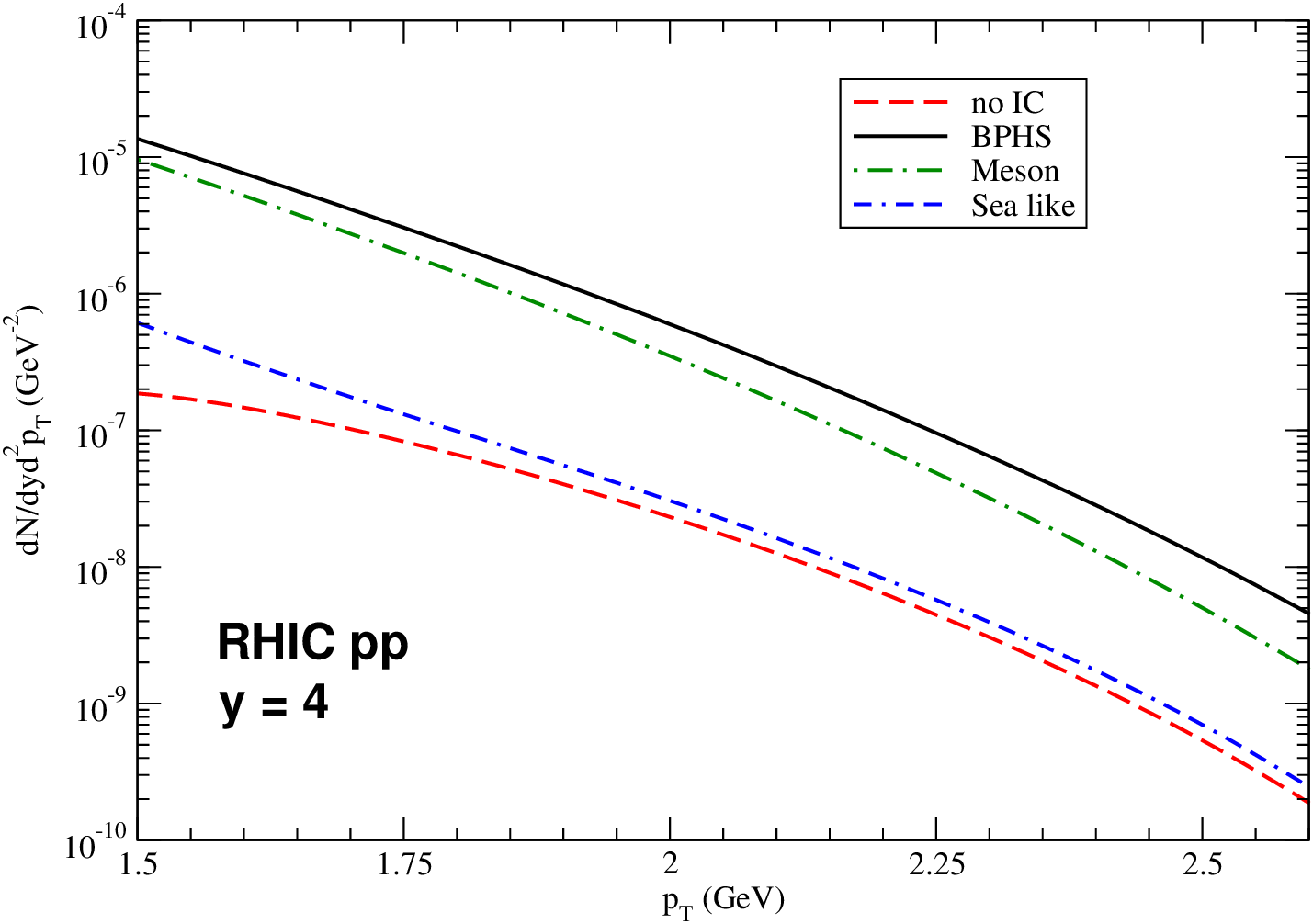} \\
\,\,\,\\
\includegraphics[scale=0.45]{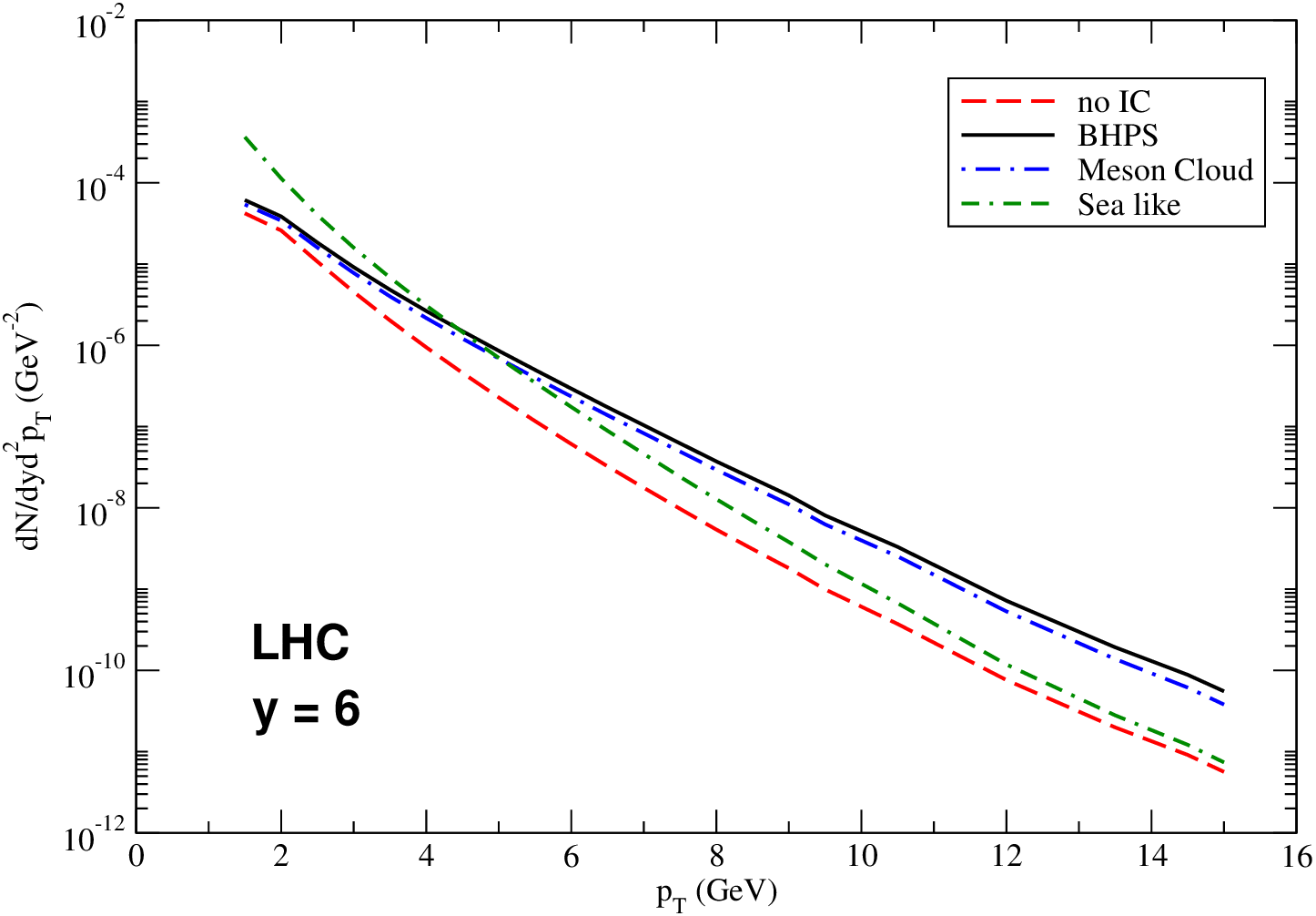}
\end{tabular}
\caption{Transverse momentum distribution of $D$ mesons produced in p-p collisions at 
RHIC and LHC. The spectra are calculated with (\ref{secdif}) 
and the four charm densities taken from \cite{cteq} and shown in Fig. \ref{fig1}.}
\label{fig2}
\end{figure}

 As the charm densities and fragmentation functions come directly from \cite{cteq} and \cite{bkk}, respectively, and the dipole scattering amplitude  has its 
parameters already fixed in \cite{buw}, the calculation of the Eq. (\ref{secdif}) is 
straightforward and  parameter free. In Fig. \ref{fig2} (upper  panel) we show the $p_T$ 
distribution of $D$ mesons predicted for proton-proton collisions at RHIC energy 
($\sqrt{s} = 200$ GeV) and fixed rapidity ($y =4$). One can see that the presence of  
intrinsic charm modifies  the magnitude of the cross section, as well as the shape of the 
$p_T$ distribution. The behavior  of the different models can be easily understood  
if we remember that:
\begin{equation}
x_p \approx \frac{p_T}{\sqrt{s}} e^y
\label{xp}
\end{equation}
From this expression we  see that, in the $p_T$ range shown in the figure, 
at RHIC energies and $y = 4$,  the typical values of $x_p$ are in the range 
$0.4 \lesssim       x_p   \lesssim 0.95$. 
In this range, all IC models predict a larger charm component than the no-IC model as shown in 
Fig. \ref{fig1}. This also implies  larger values for the cross section calculated 
with BHPS and meson cloud IC models. The cross section calculated with the sea like IC 
model is smaller, since  it predicts, already at $x \approx 0.6$, a very small charm 
distribution, quite similar to the no-IC one. On the other hand, the BHPS and meson cloud 
models predict a very large  IC distribution  in this kinematical range, which implies 
larger values for the cross section. As expected from Fig. \ref{fig1}, the BHPS prediction  
for the spectrum is larger than the meson cloud one.
Our predictions  for $D$-meson spectra at LHC energy ($\sqrt{s} = 14000$ GeV) and fixed 
rapidity ($y=6$) are shown  in Fig. \ref{fig2} (lower  panel). In this case, the typical values of 
$x_p$ are in the range: 
$0.04 \lesssim       x_p   \lesssim 0.6$, which is almost complementary to the RHIC range.
In this range the sea like model dominates at small $x_p$, while the BHPS and meson cloud 
dominate  at large $x_p$. These features are directly reflected in  Fig. \ref{fig2}
(lower panel).

\begin{figure}[t]
\begin{tabular}{c}
\includegraphics[scale=0.45] {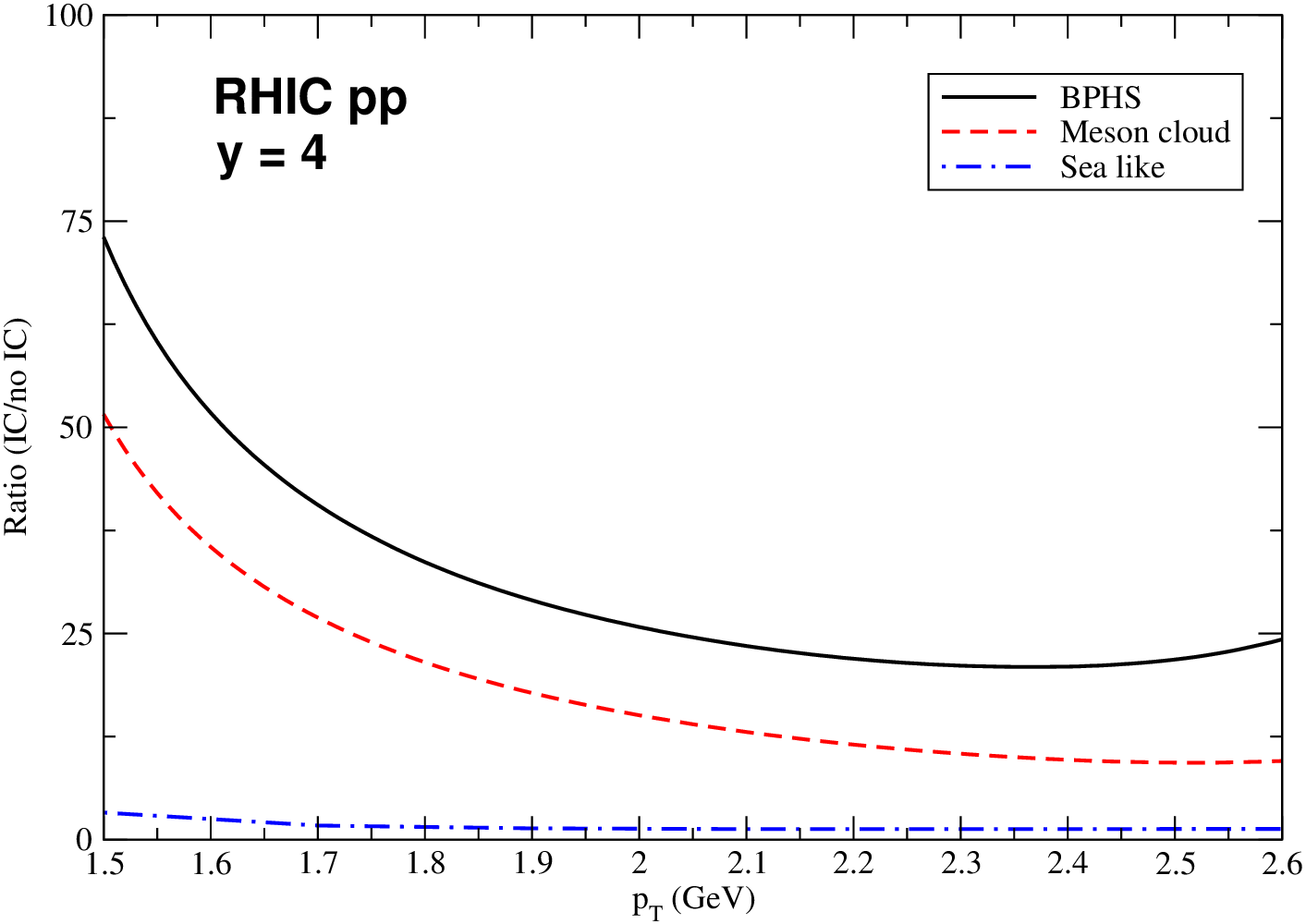} \\
\,\,\,\\
\includegraphics[scale=0.45]{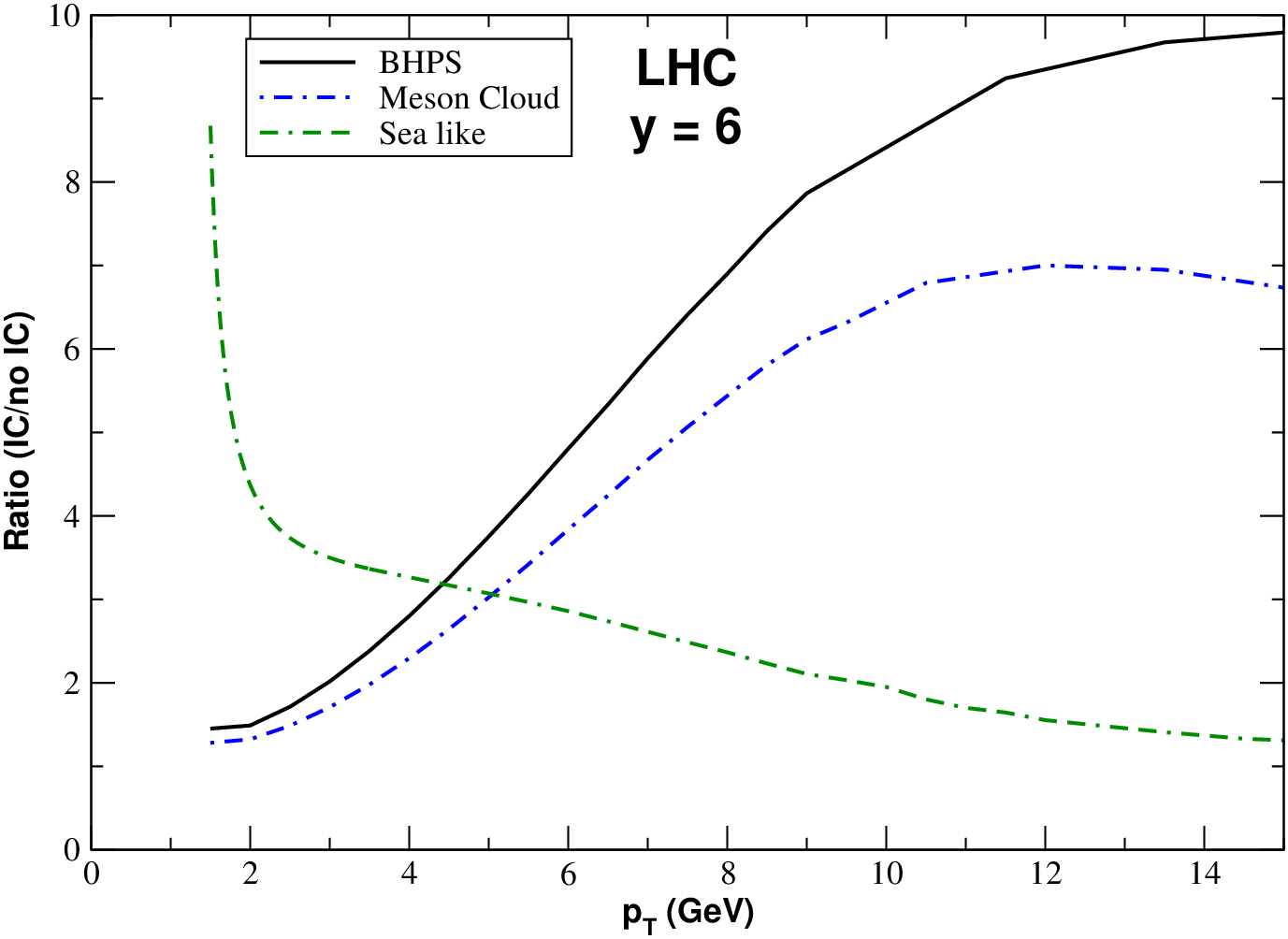}
\end{tabular}
\caption{Ratio between the different spectra calculated assuming IC and the no-IC 
spectrum shown in Fig. \ref{fig2}.}
\label{fig3}
\end{figure}

\begin{figure}[t]
\begin{tabular}{c}
\includegraphics[scale=0.45] {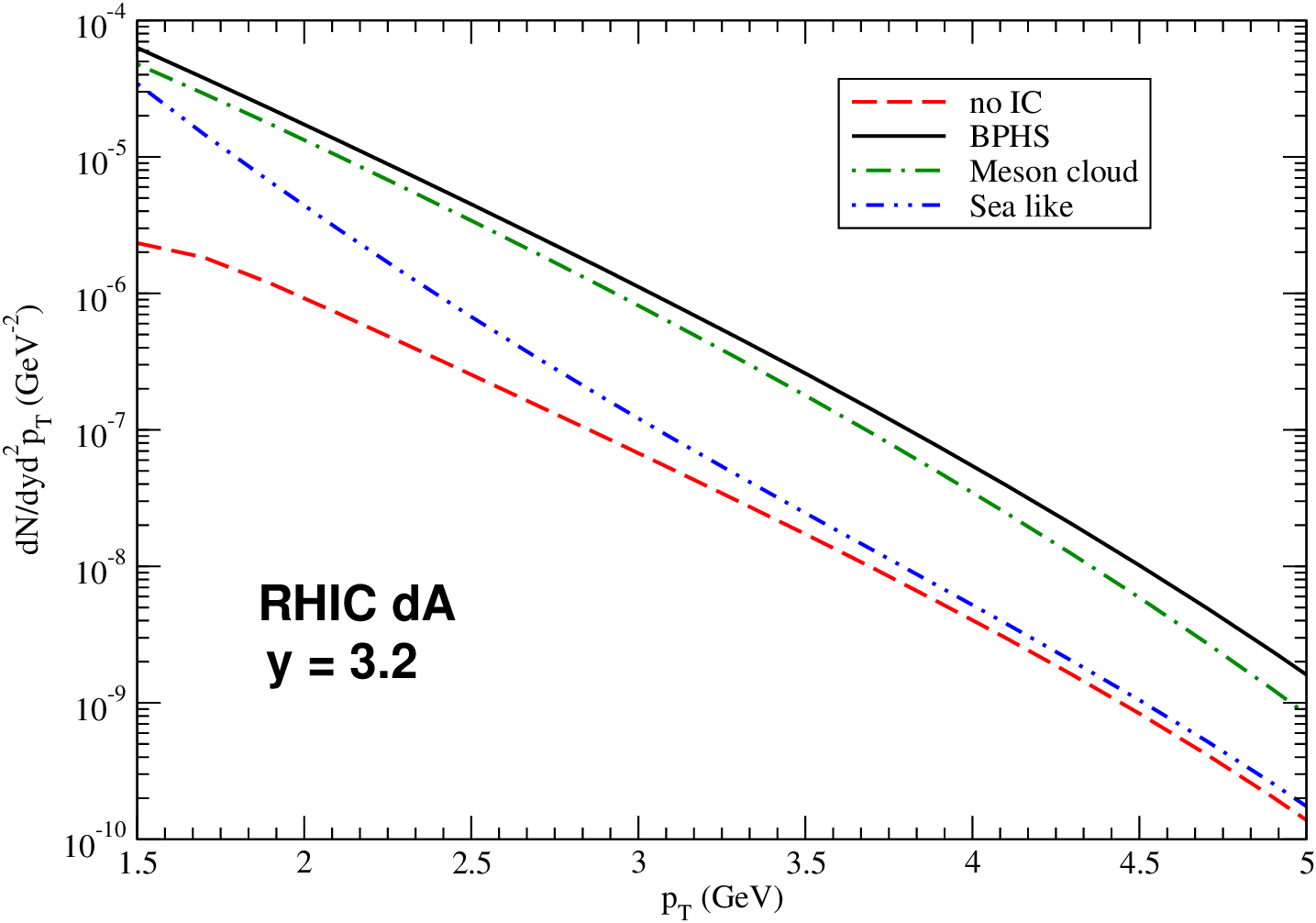} \\
\,\,\,\\
\includegraphics[scale=0.45]{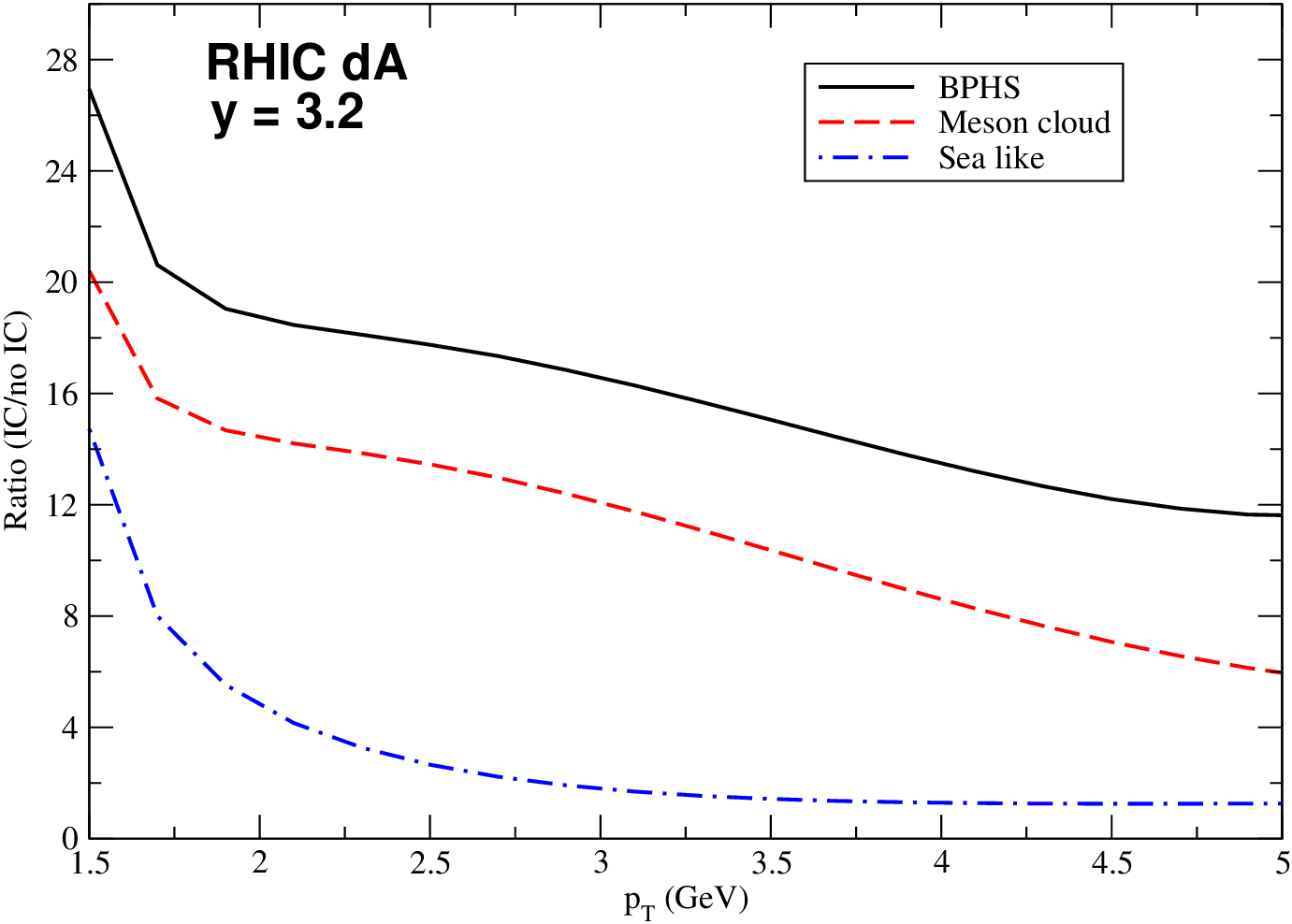}
\end{tabular}
\caption{Upper panel: spectra calculated for deuteron-gold collisions with several IC 
models and with no IC. Lower panel: ratio IC/ no IC.}
\label{fig4}
\end{figure}

In order to quantify the magnitude of the enhancement associated to the presence of  
intrinsic charm, we have estimated the ratio between the spectra predicted by the 
different IC models and the prediction without the  IC component (no IC).  In 
Fig. \ref{fig3}    we present our results for this ratio at RHIC (upper panel) and 
LHC (lower panel),  respectively. At RHIC, the enhancement predicted by the BHPS and the 
meson cloud models 
is large in the whole $p_T$ range.   On the other hand, at LHC the enhancement grows 
with $p_T$ and can be of a factor 10 at large values of  transverse momenta. The main 
advantage of using formula (\ref{secdif}) is that is able to account for the non-linear 
physics, which is expected to be more visible with nuclear targets. Therefore we shall  
extend our calculations to forward ($y=3.2$) charm production at RHIC in d-Au collisions. 
This is shown in  Fig. \ref{fig4}. Comparing with the proton-proton case, we see that in  
d-Au there is a similar enhancement in the $p_T$ spectra and it happens at lower rapidities, 
becoming easier to be observed.

\begin{figure}[t]
\begin{tabular}{c}
\includegraphics[scale=0.45] {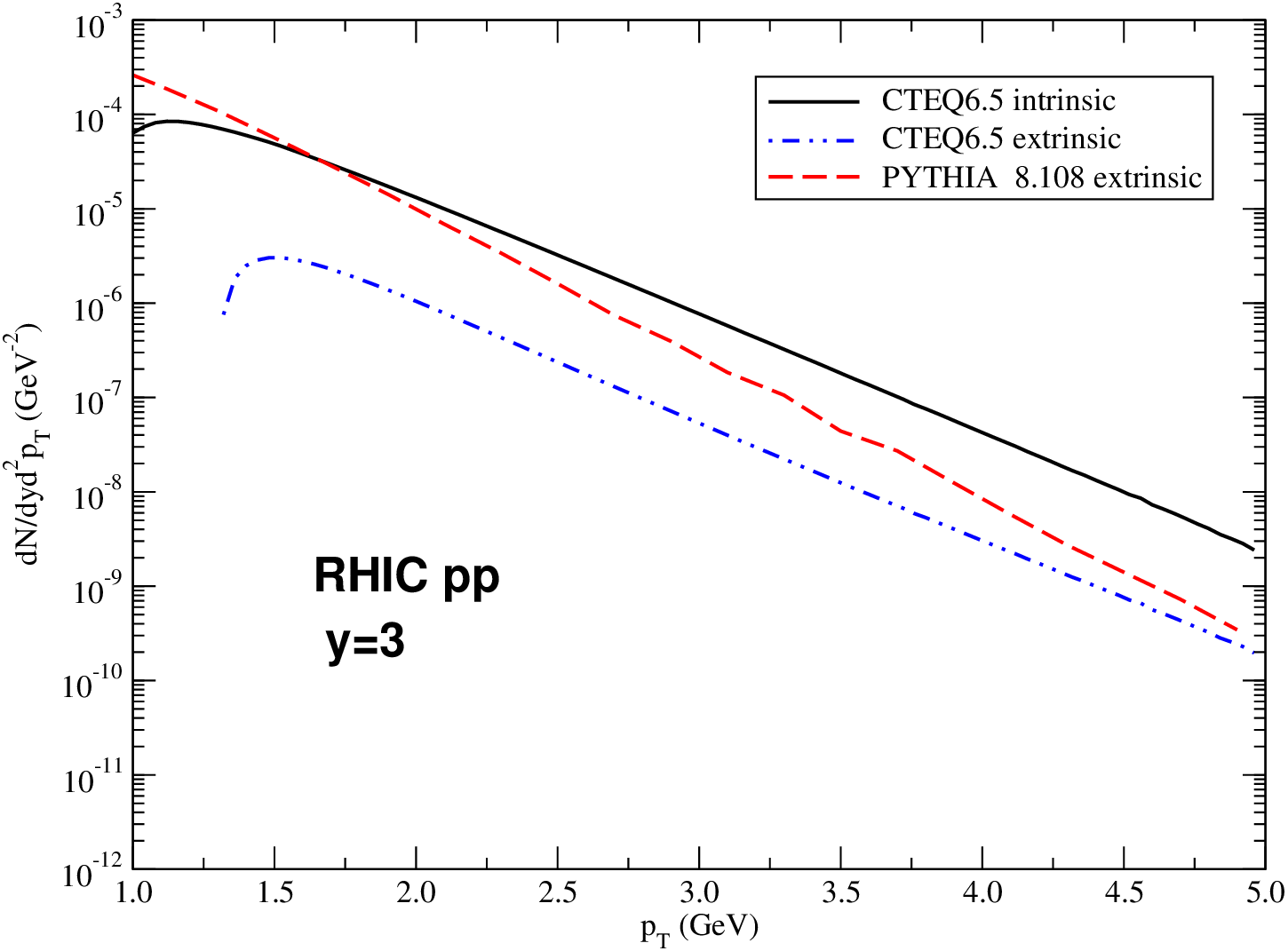} \\
\,\,\,\\
\includegraphics[scale=0.45]{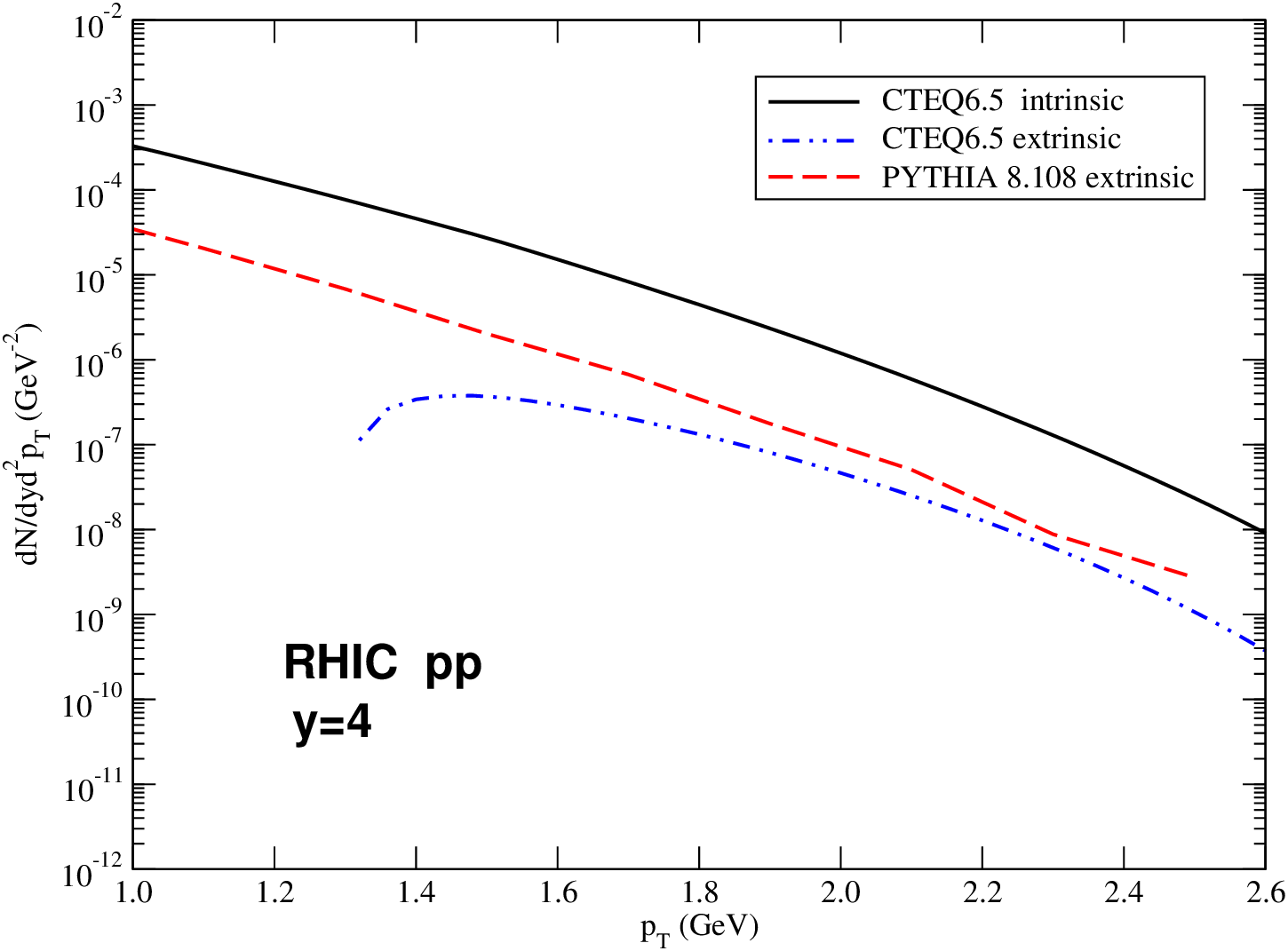}
\end{tabular}
\caption{Comparison between our spectra calculated assuming IC and the predictions of 
PYTHIA.}
\label{fig5}
\end{figure}

So far we have compared IC  with no IC predictions, with all the distributions coming 
from the same set (CTEQ) used as input for the same formula (\ref{secdif}). This is the  
correct procedure to  establish the order of magnitude of the effect that we are 
investigating. However,  there are other $D$ meson production mechanisms that contribute to 
the experimental background,  above which we are looking for IC. As mentioned above, 
charm production at high energies is described by gluon
fusion  and light quark-antiquark fusion. The elementary cross section is folded with the 
corresponding parton distributions. This is the basic procedure in most of the calculations 
performed with PYTHIA. This is the dominant production mechanism in the  central  
rapidity region. In the forward region it underpredicts the experimental distributions. 
In Fig. \ref{fig5} we compare our $p_T$ spectra obtained with 
(\ref{secdif}) and the CTEQ pdf's for the BHPS model (solid line) and no-IC (dashed line) with  the predictions of (LO) PYTHIA 8.108, supplemented 
with MRSTMCal (LO) pdf's (dot-dot-dashed curve) \cite{thorne}. We have chosen this particular implementation of 
PYTHIA because it was tuned to reproduce charm production in the central region measured 
by STAR.  
It should be mentioned that the MRSTMCal gluon distribution is a factor two larger than the 
CTEQ one in the same range of $x$ and $Q^2$ and therefore the PYTHIA curves presented in 
Fig. \ref{fig5}, in dashed lines, should be regarded as upper limits. We can observe that 
for $y=3$ (upper panel) the IC curve is comparable with PYTHIA prediction at lower values 
of $p_T$. For larger values of $p_T$ the IC curve is almost one order of magnitude larger 
than the PYTHIA estimate. For  $y=4$ (lower panel) the IC distribution is almost a factor 
$10$ larger than the  PYTHIA one in all the $p_T$ range considered. Of course, at this stage 
there are still many uncertainties in this comparison and no definite conclusion can yet be 
drawn  from Fig. \ref{fig5}. However our results are encouraging. They strongly suggest that 
the future measurement of  open charm forward $p_T$ spectra may show an enhancement over 
standard PYTHIA predictions, which is very hard (if at all!) to explain without intrinsic 
charm.

Some comments are in order here. Firstly, an alternative probe of the IC may  
be the study of the $p_T$ dependence of the  ratio ($D/\pi$) between the $D$  and $\pi$ 
production cross sections at a fixed rapidity. We  expect an enhancement of this ratio, 
as well as a different $p_T$ behavior, if  intrinsic charm is present. Secondly, although 
we  have only considered positive rapidities,  the $p_T$ spectra are identical for negative 
rapidities in the case of $pp$ collisions, i.e. our predictions depend only of $|y|$. 
However, for $pA$ collisions, this symmetry is broken, due to the different magnitudes of 
the high density effects in the proton and in the nuclei at the same $x$. As it is well 
known, these effects are amplified by  nuclei \cite{cgc}. Interestingly, the study of the 
$D$ meson production in the nucleus fragmentation region can be  useful to determine the 
presence of  IC and its behavior in a nuclei. Finally, it is important to emphasize that the study of $D$-meson production at forward rapidities as a probe of the intrinsic charm component was proposed and studied many years ago in Refs. \cite{hal,ingelman,vb}. However, these authors considered that the intrinsic charm is put on shell and emerge in a real charmed particle through soft non-perturbative interactions. The charmed meson in the final state is generated by the coalescence of the intrinsic charm with co-moving light spectator quarks. Basically, the process of $D$-meson production at forward rapidities, associated to the presence of  intrinsic charm component, proposed in these references is essentially non-perturbative. In contrast, the mechanism proposed for the first time in this paper is based on a perturbative treatment of the interaction of the intrinsic charm in the projectile with the Color Glass Condensate in the target. 

During the revision of this work, a closely related paper appeared \cite{kkss}, proposing  
the search for IC effects in  $p_T$ distributions of $D$ mesons produced in 
proton - antiproton collisions at Tevatron and measured by the CDF collaboration. In this 
kinematical set-up one can also have access to the large  $x$ component of the projectile wave 
function. Indeed, recalling (\ref{xp}), we observe that large $x_p$ is dominant in  the 
production of $D$'s either with large $y$ and moderate $p_T$, which is the region explored  
here, or with large $p_T$ and low $y$, which is the region considered in   \cite{kkss}. 
Both works reach similar conclusions and  may be regarded as complementary.

\section{Summary}
\label{sec5}
Although the direct measurements of heavy flavors in DIS are consistent with a perturbative  
origin, these experiments are not sensitive to heavy quarks at large $x$. Therefore, it is  
fundamental to study other observables which may be used to determine the presence (or not)  
of an intrinsic heavy quark component in the hadron wave function. In this paper we have  
proposed a new mechanism for the production of $D$  mesons at forward rapidities which 
assumes that the appropriate  formalism  in this kinematical range is based on CGC physics,  
and demonstrate that the  resulting transverse momentum spectra are strongly dependent on the  
behavior of the charm distribution at large Bjorken $x$. Our results indicate  that the   
intrinsic charm hypothesis can be tested in $pp$ ($p(d) A$) collisions at RHIC and LHC.

\section*{Acknowledgments}
We are deeply grateful to  Raju Venugopalan, Wally Melnitchouk, Dmitri Kharzeev and 
Gregory Soyez for fruitful discussions.   This work was  partially financed 
by the Brazilian funding agencies CNPq, FAPESP and FAPERGS.












\newpage

\newpage

\newpage

\newpage

\begin{thebibliography}{00}



\bibitem{cgc}
  E.~Iancu and R.~Venugopalan,  arXiv:hep-ph/0303204; H.~Weigert,
  Prog.\ Part.\ Nucl.\ Phys.\  {\bf 55}  (2005) 461;
  J.~Jalilian-Marian and Y.~V.~Kovchegov,
  Prog.\ Part.\ Nucl.\ Phys.\  {\bf 56}  (2006) 104.

\bibitem{forte}    F.~Caola and S.~Forte,
  Phys.\ Rev.\ Lett.\  {\bf 101}  (2008) 022001.

\bibitem{dglap} V.N. Gribov and L.N. Lipatov, Sov. J. Nucl. Phys. {\bf 15} (1972) 438;
G. Altarelli and G. Parisi, Nucl. Phys.  {\bf B126}  (1977) 298;
Yu.L. Dokshitzer, Sov. Phys. JETP {\bf 46} (1977) 641.


\bibitem{tung}
  W.~K.~Tung {\it et al.},
  JHEP {\bf 0702}  (2007) 053. 

\bibitem{cteq}
 J.~Pumplin, H.~L.~Lai and W.~K.~Tung,
  Phys.\ Rev.\  D {\bf 75}   (2007)  054029.



\bibitem{ingelman}
  G.~Ingelman and M.~Thunman,
  Z.\ Phys.\  C {\bf 73} (1997)  505.

\bibitem{brod_higgs}
  S.~J.~Brodsky, B.~Kopeliovich, I.~Schmidt and J.~Soffer,
  Phys.\ Rev.\  D {\bf 73} (2006)  113005.


\bibitem{bhps}  S.~J.~Brodsky, P.~Hoyer, C.~Peterson and N.~Sakai,
                Phys.\ Lett.\  B {\bf 93} (1980)  451.



\bibitem{hal} V.~D.~Barger, F.~Halzen and W.~Y.~Keung,
  Phys.\ Rev.\  D {\bf 25}  (1982) 112.


\bibitem{vb}  R.~Vogt and S.~J.~Brodsky,
  Nucl.\ Phys.\  B {\bf 438}  (1995) 261;  {\bf 478}  (1996) 311. 




\bibitem{pumplin}
  J.~Pumplin,
  Phys.\ Rev.\  D {\bf 73}   (2006) 114015.


\bibitem{kop_fac}
  B.~Z.~Kopeliovich {\sl et al.},
  Phys.\ Rev.\  C {\bf 72}  (2005) 054606.


\bibitem{raju_fac}
  H.~Fujii, F.~Gelis and R.~Venugopalan,
  Phys.\ Rev.\ Lett.\  {\bf 95}  (2005) 162002;  Nucl.\ Phys.\  A {\bf 780}  (2006) 146.



\bibitem{BRAHMSdata}
I.~Arsene {\it et al.},
 Phys.\ Rev.\ Lett.\  {\bf 93} (2004)  242303.


\bibitem{buw}
  D.~Boer, A.~Utermann and E.~Wessels,
  Phys.\ Rev.\  D {\bf 77}  (2008)  054014. 

\bibitem{betemps}
  M.~A.~Betemps and V.~P.~Goncalves,
  JHEP {\bf 0809} (2008)  019.


\bibitem{dhj}
  A.~Dumitru, A.~Hayashigaki and J.~Jalilian-Marian,
  Nucl.\ Phys.\ A {\bf 765}  (2006) 464; Nucl.\ Phys.\ A {\bf 770}  (2006) 57.



\bibitem{bkk}
  B.~A.~Kniehl and G.~Kramer,
  Phys.\ Rev.\  D {\bf 74} (2006)  037502.



\bibitem{pnndb}  S.~Paiva, M.~Nielsen, F.~S.~Navarra, F.~O.~Duraes and L.~L.~Barz, 
                 Mod.\ Phys.\ Lett.\  A {\bf 13} (1998)  2715; 
                 F.~S.~Navarra, M.~Nielsen, C.~A.~A.~Nunes and M.~Teixeira,  
                 Phys.\ Rev.\  D {\bf 54} (1996)  842. 







\bibitem{cdnn}
  F.~Carvalho, F.~O.~Duraes, F.~S.~Navarra and M.~Nielsen,
  Phys.\ Rev.\ Lett.\  {\bf 86} (2001)  5434.

\bibitem{smallx}
  J.~R.~Andersen {\it et al.}  [Small x Collaboration],
  Eur.\ Phys.\ J.\  C {\bf 48} (2006)  53. 


\bibitem{kt_hq} D.~Kharzeev and K.~Tuchin,
  Nucl.\ Phys.\  A {\bf 735} (2004)  248.



\bibitem{hq_sat}  K.~Tuchin,
  Phys.\ Lett.\  B {\bf 593} (2004)  66;  
  Nucl.\ Phys.\  A {\bf 798} (2008)  61;  Y.~V.~Kovchegov and K.~Tuchin,
  Phys.\ Rev.\  D {\bf 74} (2006)  054014; H.~Fujii, F.~Gelis and R.~Venugopalan,
  J.\ Phys.\ G {\bf 34} (2007)  S937.




\bibitem{jamaldumi}   A.~Dumitru and J.~Jalilian-Marian,
  Phys.\ Rev.\ Lett.\  {\bf 89} (2002)  022301. 

\bibitem{thorne}   A.~Sherstnev and R.~S.~Thorne, Eur.\ Phys.\ J.\  C {\bf 55} (2008)  553.

\bibitem{kkss}   B.~A.~Kniehl, G.~Kramer, I.~Schienbein and H.~Spiesberger,
  Phys.\ Rev.\  D {\bf 79}(2009)  094009.


\end{thebibliography}
\end{document}